\documentclass[conference]{IEEEtran}
\usepackage{color}
\usepackage{cite}
\usepackage{hyperref}
\hypersetup{
     colorlinks = true,
     linkcolor = blue,
     anchorcolor = red,
     citecolor = magenta,
     filecolor = blue,
     urlcolor = black,
}

\usepackage{framed}
\definecolor{Gray}{gray}{0.9}
\definecolor{shadecolor}{gray}{0.95}
\usepackage[most]{tcolorbox}

\usepackage{booktabs}
\usepackage{multirow}
\usepackage{makecell}
\usepackage{subfigure}

\usepackage{multibib}
\newcites{reference}{References}
\newcites{appendix}{List of Literature Review}

\definecolor{tabred}{RGB}{230,36,0}%
\definecolor{tabgreen}{RGB}{0,116,21}%
\definecolor{taborange}{RGB}{250,124,30}%
\definecolor{tabbrown}{RGB}{171,70,0}%
\definecolor{tabyellow}{RGB}{251,253,169}%
\newcommand*{\vcorr}{%
  \vadjust{\vspace{-\dp\csname @arstrutbox\endcsname}}%
  \global\let\vcorr\relax
}%

\usepackage{fancyhdr} 
\begin{document}

\title{Leveraging Architectural Approaches in Web3 Applications - A DAO Perspective Focused\thanks{These authors}}

\author{\IEEEauthorblockN{Guangsheng Yu\IEEEauthorrefmark{1}, Qin Wang\IEEEauthorrefmark{1},  Tingting Bi\IEEEauthorrefmark{1}, Shiping Chen\IEEEauthorrefmark{1}, Sherry Xu\IEEEauthorrefmark{1}
}
\IEEEauthorrefmark{1}\textit{CSIRO Data61, Australia} \\
}

\maketitle

\begin{abstract}
Architectural design contexts contain a set of factors that influence software application development. Among them, \textit{\textbf{organizational}} design contexts consist of high-level company concerns and how it is structured, for example, stakeholders and development schedule, heavily impacting design considerations. Decentralized Autonomous Organization (DAO), as a vital concept in the Web3 space, is an organization constructed by automatically executed rules such as via smart contracts, holding features of the permissionless committee, transparent proposals, and fair contribution by stakeholders. In this work, we conduct a systematic literature review to summarize how DAO is structured as well as explore its benefits\&challenges in Web3 applications. 

\end{abstract}


\section{Introduction}
Architectural design contexts influence software applications design in many ways, such as \textit{developmental}, \textit{technological}, \textit{business}, \textit{organizational}, \textit{operational}, \textit{social}, and other influences \cite{bi2018architecture}\cite{harper2015exploring}. A system of similar functionalities can work differently in different contexts \cite{petersen2009context}. Whist design contexts are important to making design decisions, the contexts of a system are often ignored and some of the design contexts may not be explicitly captured in the requirement documents \cite{bedjeti2017modeling}. \textbf{\textit{Organizational}} context, as one of the important design contexts, comprises considerations such as how a company is organized, stakeholders (e.g., developers, architects), development schedule, and financial factors \cite{bi2018architecture}. In addition, organizational design contexts are vital in many domains, which heavily impacts a set of design considerations \cite{wijerathna2022mining}. We develop the paper from this view.

The Decentralized Autonomous Organization (DAO) is an emerging term in the Web3 space, which is an important consideration of organizational design contexts. Cryptocurrency communities fall into fanaticism on DAOs due to their super decentralization and self-governance \cite{wang2022exploring}. Formed organizations replace the traditional third party with on-chain smart contracts, aiming to establish a legal structure without central authorities that enable members to seize the best interest of the entity. DAOs achieve fairness as anyone only needs to trust the transparently operating code on-chain \cite{sheridan2022web3}.

However, it is also realized that the current DAOs are still far from mature \cite{critique}. There is a set of issues existing, for example:

\begin{itemize}
\item The very beginning recognition of DAO for current participants is even absent. People are unclear about the basic concepts and definitions of DAO \cite{arroyo2022dao}.  
\item Th claimed properties given by DAOs are debatable. It is unknown to the community whether a DAO can well hold the properties such as self-governance and complete decentralization \cite{dounas2022architecture}. 
\item The potential applications of DAO lack exploitation. Whether a DAO can be widely adopted in the future is still a question. Meanwhile, external tools/protocols surrounding DAOs require further development. Participating members will definitely expect a smooth transition and better user experience from traditional organizations to this new form. 
\end{itemize}

Given those discussed issues, the entire view of DAO needs to be structurally and systematically stated. Further technical, political, and economic challenges are also required for deep exploration. In this paper, we focus on summarizing how DAO is structured in Web3 and relevant benefits/challenges from previous literature.  The main goal and three initial RQs that we try to answer are listed below:

\noindent\textbf{Research goals.} In this paper, we aim to systematically summarize \textit{the architectural contexts (in particular, DAOs) in Web3 applications, their design benefits and challenges, and the potential development directions.} We break this main goal into three research questions as follows.

\begin{itemize}
    \item \textbf{RQ-1: What are general considerations about DAOs? How a DAO is structured in Web3 applications? } Problem space exploration is important, the more structured the problem space is, the more rationally the approach can be taken by researchers and practitioners. Design considerations are raised to address design concerns, which is one of the important design activities. This RQ helps to explore the common and general considerations that researchers seek to resolve in their work. 
    \item \textbf{RQ-2: What benefits and challenges have been reported towards leveraging architectural solutions for DAOs in Web3 applications? } DAO in Web3 applications will lead to certain benefits as well as costs. The answer to this RQ can help researchers and practitioners understand the benefits and limitations of the current structured DAO in Web3 applications. 
    \item \textbf{RQ-3: What is the potential direction and development in the near future? } DAO in Web3 applications has been learned towards a better format with many state-of-the-art features. The answer to this RQ points out the development direction for researchers and practitioners.
\end{itemize}

To answer the RQs, we perform a systematic literature review that complies with Kitchenham's standard guideline~\cite{keele2007guidelines}. The objectives are to (i) provide an overview of the research activities and topics in architecture design for DAOs; (ii) understand challenges and investigate corresponding possible solutions for DAO practitioners; and (iii) point out the potential development directions for DAO designs. The key \textit{\textbf{contributions}} of this paper are as follows:
\begin{itemize}
\item We provide a comprehensive qualitative and quantitative understanding of DAO in Web3 applications, including what design considerations researchers hold and what application goals developers expect.
\item We summarize the benefits and challenges of DAOs from empirical evidence that researchers report in their work. We also list the ways of leveraging architectural designs for DAO in Web3 at present, as well as the potential development directions for DAO design.
\end{itemize}
\label{Sec_Intro}

\section{Research Design} 
\subsection{Research Methodology}

This literature review consists of three stages: (i) \textbf{planning}: before conducting the literature review, we prepared a research protocol during the planning phase that focuses on the specific objectives of understanding DAO in Web3 applications; (ii) \textbf{literature study selection and analysis execution} is described in details in Sec.\ref{subsec_research_process}. and (iii) \textbf{report the results}: we report and discuss our results in Sec.\ref{sec_results}.

\subsection{Research Process}
\label{subsec_research_process}

\noindent{\textbf{Study search and selection}}. We conducted this literature review in order to understand the state of the art of DAO in Web3 applications. We defined two research questions (RQs), which concern Sec.\ref{Sec_Intro}, which concern how DAO is structured in Web3 applications and what benefits and challenges of DAO  that summarized in previous literature. We define the search scope, search strategy, and selection criteria for conducting this literature review. 

\begin{itemize}
\item \textit{Time period}: we planned to search as many research papers as possible to get a comprehensive understanding of this topic, however, as Web3 is a new concept, which has emerged within five years, so we did not define the start time to reduce risks of omitting some literature, and the end time was set on October 2022.
\item \textit{Electronic databases}: we collected the eligible candidate papers from five mainstream academic databases, covering 
$\mathsf{Science Direct}$, $\mathsf{Springer Link}$, $\mathsf{ACM Digital Library}$, $\mathsf{ISI Web of Science}$, and $\mathsf{IEEE explore}$, to retrieve related literature. $\mathsf{Google Scholar}$ was not included in this study since it will produce a number of unrelated results, and the retrieved literature overlaps with the previous five indexed databases.
\end{itemize}

\noindent\textbf{Search strategy.} Search strategy influences the quality of the retrieved literature and determines the time and effort required to search the literature. The search strategy in this mapping study was divided into two steps:
\begin{itemize}
\item We defined the search terms based on the topics, as such the keywords we selected are: ``Software Architecture" \textbf{AND}  `DAO''  \textbf{OR} ``Blockchain Governance".
\item  We define two paper selective criteria: (i) A study published in full-text and written in English; and (ii) A study that investigates DAO in Web3 application, and (iii) Keywords should be included in ``Title or Abstracts'' of the studies. We also define several exclusive criteria: (i) if a study investigates DAO but not discusses architectural design (design contexts); and (ii) if a study investigates architecture design contexts but not discusses DAO.
\end{itemize}

Finally, we collected 49 relevant literature (listed in Appendix), and we answer the RQs in Sec.\ref{sec_results}.

\begin{table}[!hbt]
    \centering
    \caption{Categories of Selected Publications}\label{tab-category}
    \resizebox{\linewidth}{!}{
    \begin{tabular}{rlcr}
    \toprule
    \multicolumn{1}{c}{{\textbf{Category}}}   & \multicolumn{1}{c}{{\textbf{Scoping}}} & \multicolumn{1}{c}{{\textbf{Count}}} &  \multicolumn{1}{c}{{\textbf{Reference}}}  \\

  \midrule
  \multirow{2}{*}{ \rotatebox{0}{\textit{{\textbf{Managerial}}}}}
   
   & Legality and Liability &  3 &  \cite{rodrigues2018law}\cite{ostbye2022exploring}\cite{minn2019towards} \\
   & Sociology and Ethics & 2 & \cite{hutten2019soft}\cite{sulkowski2019tao} \\
   
   \midrule
   
     \multirow{3}{*}{\textit{\textbf{\makecell{Framework \\ \& Design}}}} 
     & Supportive middleware and architecture &  2 & \cite{yue2020blockchain}\cite{mehdi2022data}\\
     & Self-governance, utility token, e-voting & 2 & \cite{ding2021parallel}\cite{mackey2019framework} \\
     & Applications and use cases & 6  & \cite{bischof2022longevity}\cite{qin2020blockchain}\cite{jeyasheela2021blockchain}\cite{diallo2018egov}\cite{nguyen2019leveraging}\cite{zichichi2019likestarter} \\
     
   \midrule

   \multirow{4}{*}{\textit{\textbf{\makecell{Survey \\ \& SoK}}}}
   & Self-governance, utility token, e-voting &  6 &  \cite{morrison2020dao}\cite{rikken2019governance}\cite{dirose2018comparison}\cite{reijers2021now}\cite{singh2020computational}\cite{shermin2017disrupting} \\
   & Crypto, blockchain techniques and tools & 2 & \cite{singh2019blockchain}\cite{beniiche2021way} \\
   & DAO projects on blockchain platforms & 4 & \cite{liu2021technology}\cite{faqir2021comparative}\cite{el2020overview}\cite{wang2019decentralized}\\   
  & DAO as a component in blockchain & 3 &\cite{wang2019blockchain}\cite{ding2022desci}\cite{takagi2017organizational}\\   
   
   \bottomrule
    \end{tabular}
    }
\end{table}



   

\section{Results and Discussion}
In this section, we analyze each paper in the pool and accordingly deliver the results surrounding research questions. Specifically, we discuss the aspects of \textit{definitions}, \textit{structure}, \textit{benefits}, \textit{challenges}, and \textit{future directions}.

\begin{figure}[!hbt]
    \centering
    \includegraphics[width=0.85\linewidth]{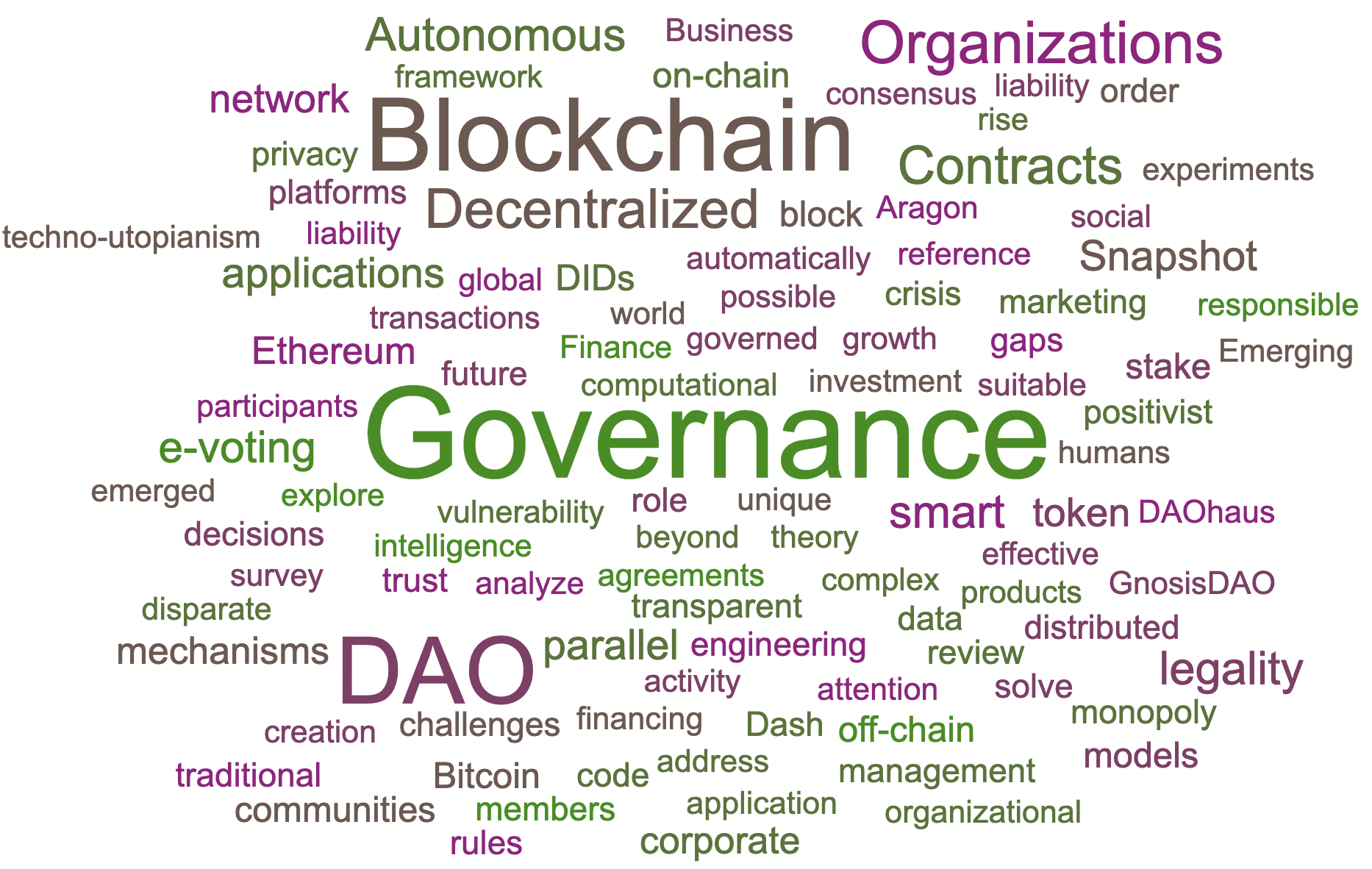}
    \caption{Answering \textit{{RQ-1: What is DAO?}} The figure is plotted based on statistical results of keyword frequency by the inputs of in-pool publications.}
    \label{fig:word_cloud}
\end{figure}

\subsection{Answering RQ-1}

To answer RQ-1, the definition of DAO reported by each study is recorded. This question helps the audience to understand both \textit{the DAO scope} and \textit{the perceptions of researchers on DAO}. Fig.\ref{fig:word_cloud} shows a word cloud of the frequency of the words that appear in DAOs' definitions in each study. The most frequently appeared words include: \textit{\textcolor{teal}{governance}}, \textit{\textcolor{teal}{blockchain}}, \textit{\textcolor{teal}{decentralized}}, \textit{\textcolor{teal}{organizations}}, \textit{\textcolor{teal}{contracts}}, \textit{\textcolor{teal}{autonomous}}, \textit{\textcolor{teal}{management}}, \textit{\textcolor{teal}{decisions}}, \textit{\textcolor{teal}{communities}}, \textit{\textcolor{teal}{corporate}}, etc. More precisely, we use four categories to classify the mentioned definitions: (i) execution workflow, (ii) data indexing, (iii) self-governance, and (iv) human-centric (see Tab.\ref{tab:rq1}).

First, in the \textit{execution workflow}, most researchers consider that all processes and operations in DAOs are executed by running decentralized smart contracts to achieve autonomous organizing and management. This can be observed as the most frequently mentioned keyword in RQ-1 is \textit{\textcolor{teal}{smart contracts (SC) for autonomous organization}}. Other frequently mentioned keywords describing the execution workflow include \textit{\textcolor{teal}{parallel}}, \textit{\textcolor{teal}{distributed}}, and \textit{\textcolor{teal}{corporate}}. Besides, \textit{\textcolor{teal}{smart contracts impacting on the decision-making}} is another characteristic that describes how the execution workflow in DAOs is performed in a decentralized manner. Smart contracts have had impacts on DAOs where decision-making is distributed or delegated away from a central authority. This also highlights how researchers differentiate DAO from traditional organizations. Authors in~\cite{wang2022empirical} also state that a widely used structure of DAO-oriented smart contracts in Web3 applications is \textit{\textcolor{teal}{multi-sig}} wallets for secure asset reservation and set voting strategies for fair governance. Smart contracts enable real-time auditing and verification, hence enhancing the machine-execution security~\cite{tolmach2021survey}.

\begin{table}[!hbt]
    \centering
    \caption{Categories of DAOs}
    \resizebox{\linewidth}{!}{
    \begin{tabular}{rlcr}
    \toprule
    \multicolumn{1}{c}{\textbf{Category}}   & \multicolumn{1}{c}{\makecell[c]{\textbf{Characteristic}}} & \multicolumn{1}{c}{\textbf{Count}} & \multicolumn{1}{c}{\textbf{Reference}} \\
    
    \midrule
    
    \multirow{2}{*}{\textit{\textbf{Execution workflow}}}
   & SC for autonomous organization & 5 & \cite{yue2020blockchain}\cite{ding2021parallel}\cite{bischof2022longevity}\cite{qin2020blockchain}\cite{ding2022desci} \\
   & SC impacting on decision-making & 2 &  \cite{qin2020blockchain}\cite{shermin2017disrupting}  \\
   
   \cmidrule{2-2}
    \multirow{1}{*}{\text{\text{Applications}}}
   & Multi-sig wallets &  2 & \cite{wang2022empirical}\cite{beniiche2021way} \\

   \midrule
   
    \multirow{2}{*}{\textit{\textbf{Data indexing}}}
   &  On-chain identifier  &  3 & \cite{w3cdid}\cite{weyl2022decentralized}\cite{reijers2021now} \\
   &  Off-chain snapshot & 2 & \cite{wang2022empirical}\cite{reijers2021now}\\
   
   \cmidrule{2-2}
    \multirow{2}{*}{\text{\text{Applications}}}
   & Decentralized identifiers (DIDs) & 3  &   \cite{wang2022empirical}\cite{w3cdid}\cite{weyl2022decentralized} \\
   & Snapshot &  1 & \cite{wang2022empirical} \\
   
   \midrule
   
    \multirow{2}{*}{\textit{\textbf{Self-governance}}}
       & By means of e-voting &  \multicolumn{2}{c}{9 \cite{minn2019towards}\cite{mehdi2022data}\cite{ding2021parallel}\cite{mackey2019framework}\cite{morrison2020dao}\cite{rikken2019governance}\cite{dirose2018comparison}\cite{reijers2021now}\cite{singh2020computational}} \\
      & Stake and utility token   & 2 & \cite{reijers2021now}\cite{faqir2021comparative}  \\

   \cmidrule{2-2}
   \multirow{2}{*}{\text{\text{Applications}}}
   & Consensus procedures & 2  & \cite{singh2019blockchain}\cite{wang2019decentralized} \\
   & Defending against Sybil attacks &  2 & \cite{wang2019decentralized}\cite{ding2022desci} \\    
   
   \midrule
   
    \multirow{2}{*}{\textit{\textbf{Human-centric}}}
       & Liability, legality and ethics & 5 & \cite{rodrigues2018law}\cite{ostbye2022exploring}\cite{hutten2019soft}\cite{sulkowski2019tao}\cite{morrison2020dao}  \\
   & Social, human, communities   & 3 & \cite{liu2021technology}\cite{el2020overview}\cite{wang2019blockchain} \\

   \cmidrule{2-2}
   \multirow{2}{*}{\text{\text{Applications}}}
   & Finance and Crowdfunding  &  4 & \cite{bischof2022longevity}\cite{jeyasheela2021blockchain} \cite{zichichi2019likestarter}\cite{fritsch2022analyzing}   \\
   & Law and Democracy & 3 & \cite{rodrigues2018law}\cite{diallo2018egov}\cite{takagi2017organizational} \\

   \bottomrule
    \end{tabular}
    }
    \label{tab:rq1}
\end{table}

Secondly, DAOs can be discussed in terms of \textit{data indexing}. It is realized that researchers categorize data indexing into two sectors, i.e., on-chain and off-chain. The keywords \textit{\textcolor{teal}{on-chain identifier}} and \textit{\textcolor{teal}{off-chain snapshot}} highlight the on-chain and off-chain techniques, respectively. On-chain identifiers are used to enable globally unique, secure, and cryptographically verifiable authentication services. Off-chain snapshots are used to enable the in-time data status to improve look-up efficiency and smooth collaboration. On-chain \textit{\textcolor{teal}{decentralized identifiers}} (DIDs)~\cite{w3cdid} are used instead of traditional identifiers in the absence of any central entity. By making use of Public Key Infrastructure (PKI) technology in Web3 applications to generate asymmetric key pairs stored on-chain, DIDs can achieve globally unique, secure, and cryptographically verifiable authentication services. Typical examples of implementation of DIDs include Ethereum address and Ethereum name service (ENS) \cite{ens}. On the other hand, \textit{\textcolor{teal}{Snapshot}} is mentioned for off-chain data caching which is expected to be structured in many Web3 applications to smooth the collaboration between multiple parties and improve the efficiency of governance.

The third category is \textit{self-governance} which is the most crucial component in DAOs by \textit{\textcolor{teal}{governance}} being reflected in the word cloud the most, as shown in Fig.\ref{fig:word_cloud}. The keywords of this category found are \textit{\textcolor{teal}{by means of e-voting}} and \textit{\textcolor{teal}{stake and utility token}}. The self-governance is mainly conducted for decentralized decision-making by running e-voting across members who own a certain amount of stake or utility tokens. They are used for governance and votes, while at the same time representing an on-chain reputation or voting power of a unique DID during e-voting and being subject to \textit{\textcolor{teal}{consensus procedures}}. This brings the token system a strong capability of \textit{\textcolor{teal}{defending against Sybil attacks}}~\cite{sybil}. 

The last category is \textit{human-centric} which focuses on humans/communities and responsible activities. Many studies mention the words of \textit{\textcolor{teal}{liability}}, \textit{\textcolor{teal}{legality}}, and \textit{\textcolor{teal}{ethics}}. These keywords delineate a partial landscape of DAO's vision. DAO organizations should be responsible for positive and ethical activities. Another line of keywords \textit{\textcolor{teal}{society}}, \textit{\textcolor{teal}{human}}, and \textit{\textcolor{teal}{communities}} reflects the participation scale of DAOs or their impacts. We can find that many DAOs have rooted in human-centric behaviors for building a sustainable ecosystem. Finally, as a summary, by connecting and grouping each keyword with others under the same definition, a picture of answering RQ-1 can be concluded and summarized below.

\begin{center}
\tcbset{
        enhanced,
        boxrule=0.4pt,
        fonttitle=\bfseries
       }
\begin{tcolorbox}[colback=gray!10,
colframe=black, 
width=3.4in,
arc = 1mm,
boxrule=0.5 pt,
lifted shadow={1mm}{-2mm}{3mm}{0.1mm}{black!50!white}
]
\footnotesize
\textbf{Findings of RQ-1: \textit{What is DAOs and How a DAO is structure in Web3 applications?}}
\vspace{0.2em}

\noindent\hangindent 0.6em$\triangleright$\,\textbf{DAO.} DAO is an emerging term to describe self-governable organizations in a decentralized context. It relies on cryptography and contains smart contracts, on-chain identifiers, off-chain snapshots, and e-voting-based governance with stake and utility tokens.

\vspace{0.2em}
\noindent\hangindent 0.6em$\triangleright$\,\textbf{Approach to construct DAOs in Web3.} (i) Smart contracts used as \textit{multi-sig} wallets for secure asset reservation and set \textit{voting strategies} for fair governance. (ii) On-chain DIDs used to authenticate and authorize identities, also used for token storage and trading. Off-chain Snapshot is used to smooth collaboration and improve the efficiency of data look-up and self-governance. (iii) Stake and utility tokens representing voting power and reputation by relying on consensus procedures, which offer reliable defenses against Sybil attacks.
\end{tcolorbox}
\end{center}

\subsection{Answering RQ-2}
\begin{figure}[t]
\centering
\subfigure[What are the benefits?]{
\resizebox{0.42\textwidth}{!}{
    \includegraphics[width=0.95\linewidth]{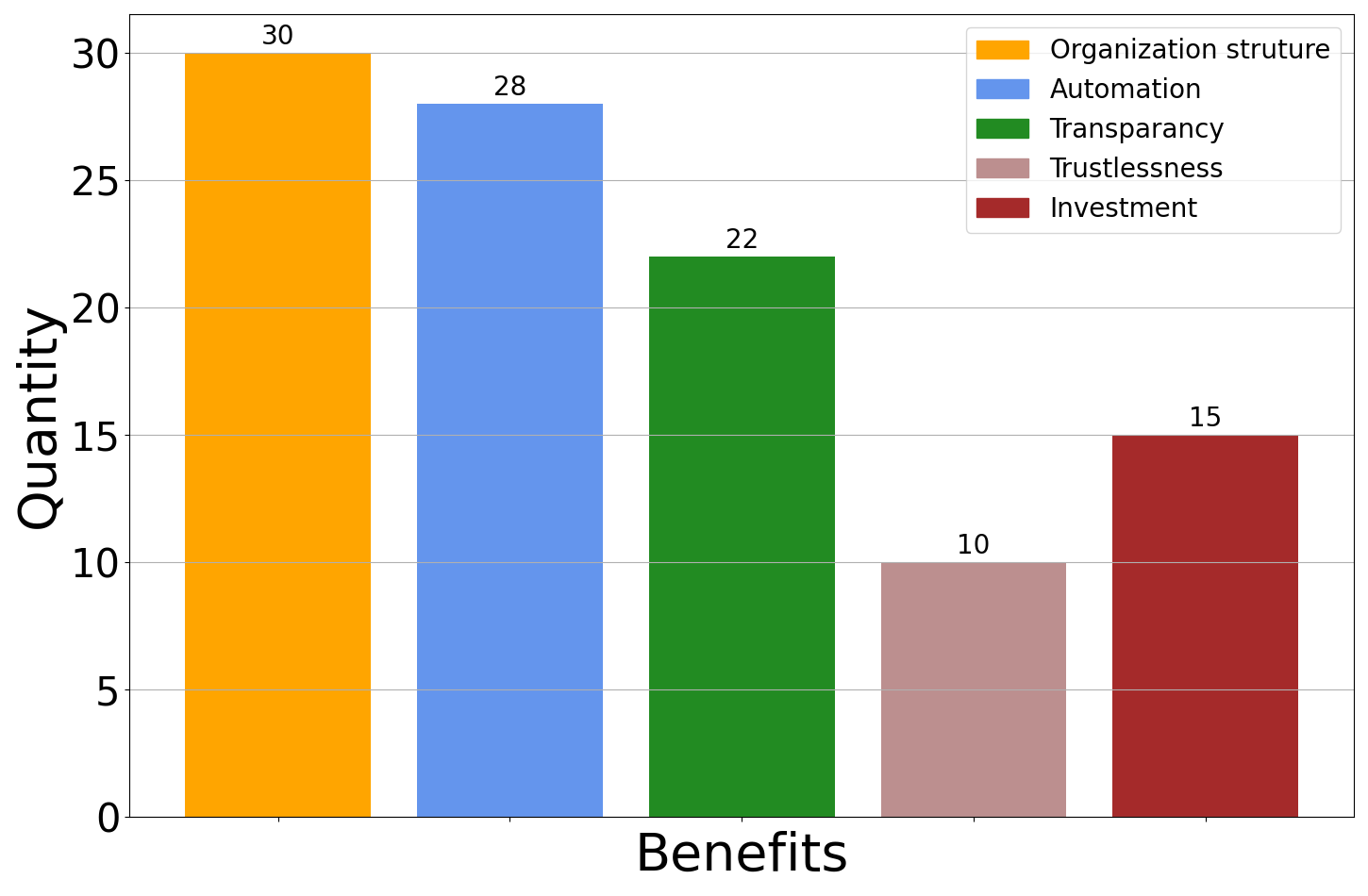}
\label{fig:benefits}
}
}

\subfigure[What are the challenges?]{
\resizebox{0.42\textwidth}{!}{
    \includegraphics[width=0.95\linewidth]{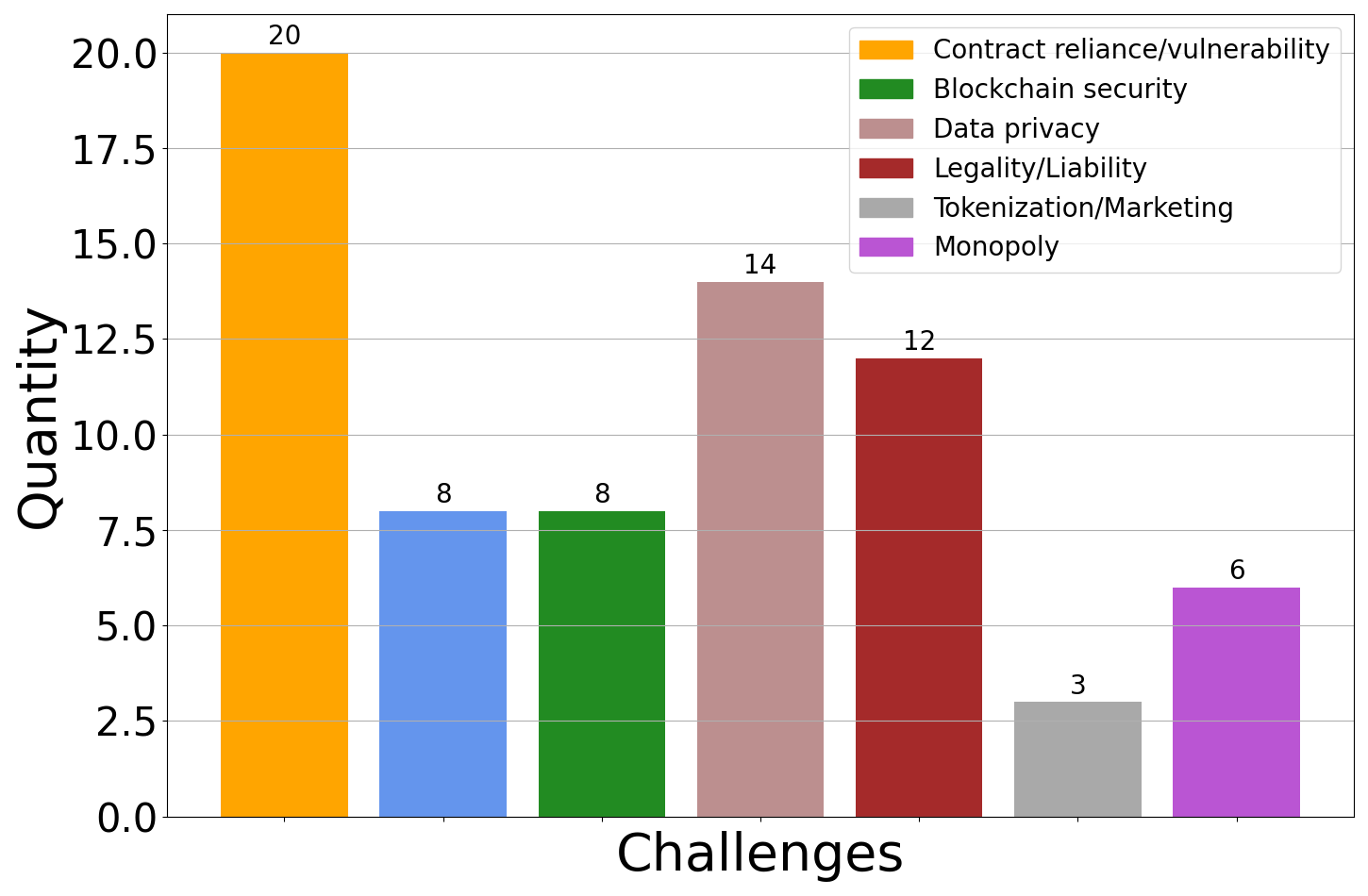}
\label{fig:challenges}
}
}
\caption{RQ-2}
\label{fig:rq2}
\end{figure}

We study the benefits and challenges of leveraging architectural solutions for DAOs through RQ-2 (cf. Fig.\ref{fig:rq2}). 

\smallskip
\noindent\textbf{Organization structure.}  
It can be realized that the most frequently mentioned benefit by researchers is the \textit{\textcolor{teal}{organization structure}}, i.e., all selected publications mention this term. The organizational structure of DAOs refers to the flat structure without relying on central entities. This also tells the most critical pain point of a traditional organization, i.e., centralization which is thought to be responsible for unexpected monopoly, manipulation, corruption, and inefficiency of management. 

\smallskip
\noindent\textbf{Automation.}
The term \textit{\textcolor{teal}{automation}} comes second (28 publications) with nearly the same level of concentration as the organization structure. Researchers consider that automated execution significantly improves the efficiency and stability of management or holding any campaigns in DAOs with non-human behaviors being involved after the development has been clearly finalized and thoroughly evaluated. 

\smallskip
\noindent\textbf{Transparency and openness.}
There are 22 publications mentioning \textit{\textcolor{teal}{transparency}} and \textit{\textcolor{teal}{openness}} in the collection. The smart contracts are open-source and are transparently executed by blockchains where every entity can verify and validate the correctness and confirm rules or policies prior to participating in any DAO campaigns. This, in fact, can create strong trustworthiness by transparent behaviors being conducted in a decentralized manner with no human manipulation involved, which also corresponds to the fourth benefit being mentioned.

\smallskip
\noindent\textbf{Trustlessness.}
The strong trustworthiness created by the factors above refers to a \textit{\textcolor{teal}{trustless}} trust, which is mentioned only 10 times in the publication pool. The trustworthiness being trustless is important in DAOs, highlighting strong system reliability in a decentralized manner.

\smallskip
\noindent\textbf{Investment.}
Researchers also take interest in the positive financial \textit{\textcolor{teal}{investment}} return in DAOs. This refers to a healthy tokenization \cite{wang2022empirical} where investors can reap profits owing to their rising capital in an influential DAO in which the price of the stake and utility tokens would accordingly increase.


\begin{center}
\tcbset{
        enhanced,
        boxrule=0.4pt,
        fonttitle=\bfseries
       }
\begin{tcolorbox}[colback=gray!10,
colframe=black, 
width=3.4in,
arc = 1mm,
boxrule=0.5 pt,
lifted shadow={1mm}{-2mm}{3mm}{0.1mm}{black!50!white}
]
\footnotesize
\textbf{Findings of RQ-2.1: \textit{What benefits have been reported towards leveraging architectural solutions for DAOs?}}
\vspace{0.2em}

\noindent\hangindent 0.6em$\triangleright$\,\textbf{Benefits of architectural DAOs.} Flat organization structure and automated execution are the two main benefits. A moderate number of studies pay attention to transparency and positive investment return, while only a small number of studies mention trustlessness. This may give a reminder of the need to emphasize in future studies how the trust is established and how the trust works under the hood in DAO spaces. 

\end{tcolorbox}
\end{center}

\smallskip
\noindent\textbf{Contract reliance/vulnerability.}
A large number of works (20 publications) in the collection mentioned the excessive \textit{\textcolor{teal}{reliance of smart contracts}} could be risky due to the hidden vulnerabilities or bugs in current smart contracts. All of the publications highlight this factor by also mentioning the huge implication of \textit{The DAO hack} that happened in 2016. Automated execution of smart contracts running on distributed nodes increases the difficulty of risk management after an attack is successfully leveraged in current DAOs.

\smallskip
\noindent\textbf{Blockchain security.}
Beyond the contract component, there are 8 publications diving into blockchain technology to figure out the security issues of blockchain that are inherited by DAOs. Researchers take into account the use of DIDs and discuss the threats of the Sybil attack and selfish attack~\cite{selfish} which are the two main attacks being noticed in publications.

\smallskip
\noindent\textbf{Data privacy.}
\textit{\textcolor{teal}{Data privacy}} is also a factor being discussed by 8 publications. Transparency and openness are beneficial for public auditing and encourage the creation of trustlessness. However, this does not mean privatizing data would not become a crucial requirement in DAOs. For example, healthcare data across hospitals or private facilities that intend to be organized via a DAO may not be willing to share the data. Current DAOs lack enough concentration on this factor.

\smallskip
\noindent\textbf{Legality/Liability.}
Half of the studies (12 publications) mention that state-of-the-art DAOs still await \textit{\textcolor{teal}{legislation}} to consolidate the business rules of DAOs. DAOs have not been recognized as legal entities in most countries. Participants cannot make a profit via organizations with corporate personhood. The absence of legal recognition potentially increases the burden on individual members as they need to be responsible for both their personal and organizational liabilities. 

\smallskip
\noindent\textbf{Tokenization/Marketing.}
Upon the legislation, investing DAOs could become a long-term strategy (mentioned by 3 publications) in a more healthy \textit{\textcolor{teal}{tokenization}} ecosystem and non-opportunistic crypto-market. This would also encourage marketing and advertisement towards those who have not yet been Web3 fans. A stable space could provide huge business opportunities that could not be resisted.

\smallskip
\noindent\textbf{Monopoly.}
Researchers start to realize that the abuse of stakes or utility tokens in e-voting processes may cause \textit{\textcolor{teal}{monopoly}}. The governance in DAOs relies prominently on the possession of stakes and utility tokens. Although it is originally expected to be the core of the decentralization in DAOs, highly active groups of participants are likely to accumulate major shares of tokens (a.k.a., Matthew effect), hence breaching the decentralization due to the concentration of e-voting power.


\begin{center}
\tcbset{
        enhanced,
        boxrule=0.4pt,
        fonttitle=\bfseries
       }
\begin{tcolorbox}[colback=gray!10,
colframe=black, 
width=3.4in,
arc = 1mm,
boxrule=0.5 pt,
lifted shadow={1mm}{-2mm}{3mm}{0.1mm}{black!50!white}
]
\footnotesize
\textbf{Findings of RQ-2.2: \textit{What challenges have been reported towards leveraging architectural solutions for DAOs?}}
\vspace{0.2em}

\noindent\hangindent 0.6em$\triangleright$\,\textbf{Challenges of architectural DAOs.} Most of the known challenges of leveraging architectural DAOs appear to be corresponding to the known benefits, including the difficulty to manage risks due to the decentralized structure and automated execution, difficulty to ensure data privacy in a transparent and open context, a management risk of monopoly due to the token accumulation in trustless e-voting processes, and financial risk of investment and marketing in the absence of legislation. 
\end{tcolorbox}
\end{center}

\subsection{Answering RQ-3}
We study the potential development directions for DAOs through RQ-3 (cf. Tab.\ref{tab:rq3}). 

\begin{table}[!hbt]
    \centering
    \caption{Categories of Potential Directions for DAOs}\label{tab:rq3}
    \resizebox{\linewidth}{!}{
    \begin{tabular}{rlcr}
    \toprule
    \multicolumn{1}{c}{{\textbf{Count}}} &  \multicolumn{1}{c}{{\textbf{Reference}}} &  
    \multicolumn{1}{c}{{\textbf{Category}}}   & \multicolumn{1}{c}{{\textbf{Scoping}}} \\

  \midrule
     \multirow{2}{*}{\textit{\textbf{\makecell{Collaboration \\ \& Management}}}}
   
   & SubDAOs organization  &  2 &  \cite{wang2022exploring}\cite{wang2022empirical} \\
   & Multi-DAOs collaboration & 3 & \cite{wang2022exploring}\cite{wang2022empirical}\cite{weyl2022decentralized} \\
   
   \midrule
   
     \multirow{2}{*}{\textit{\textbf{\makecell{Technical Tools}}}} 
     & Programming tools & 4 &  \cite{wang2022exploring}\cite{wang2022empirical}\cite{singh2019blockchain}\cite{beniiche2021way}  \\
     & Privacy-preserving approaches &4 &  \cite{wang2022exploring}\cite{wang2022empirical}\cite{yue2020blockchain}\cite{mehdi2022data} \\

   \bottomrule
    \end{tabular}
    }
\end{table}

\smallskip
\noindent\textbf{Organization of sub-DAOs.} Sub-DAOs are an emerging approach for different working groups to create their own foundation and ownership structure, which is also mentioned in 2 studies. Organizing sub-DAOs within a larger DAO can be typically managed in several ways, including (i) \textit{function-based}: sub-DAOs are organized according to their specific functions or areas of responsibility, such as finance, marketing, or operations \cite{wang2022empirical}; (ii) \textit{hierarchical}: sub-DAOs are structured in a hierarchical way in which each sub-DAO need to report to a higher-level sub-DAO or the parent DAO; (iii) \textit{geographic}: sub-DAOs are managed based on geographic regions, allowing for decentralized decision-making and localized implementation; and (iv) \textit{hybrid}: sub-DAOs are organized by a combination of any of aforementioned structures.
Similar to building management across any different groups, clear lines of communication, decision-making processes, and accountability would be established to ensure effective coordination and collaboration among sub-DAOs.

\smallskip
\noindent\textbf{Collaboration in multi-DAOs.} Collaboration in multi-DAOs can refer to the cooperation between different DAOs to achieve common goals or to achieve their individual goals in a more efficient and effective way. Unlike subDAOs, multi-DAO situations do not hold a close relationship between different DAOs. There are 3 publications mentioning the \textit{\textcolor{teal}{interaction}}, \textit{\textcolor{teal}{collaboration}} and \textit{\textcolor{teal}{competition}},  between different DAOs, as well as the design of typical \textit{\textcolor{teal}{decentralized negotiation}} protocols. Positive collaboration can be accomplished through the use of cross-DAO communication protocols, shared resources, and shared decision-making processes. Several key techniques include transparent communication, aligned incentives, and a shared governance framework. Additionally, the use of decentralized technologies like blockchain and smart contracts can help ensure that all parties can trust each other and work together effectively, even if they are geographically dispersed. Notably, at the same time, the competition among different DAOs cannot be overlooked. Competition can drive innovation and efficiency within a multi-DAO ecosystem, as each DAO strives to offer the best products, services, and solutions to its stakeholders. However, competition on the flip side can also lead to fragmentation and inefficiencies, particularly if there is a lack of coordination and collaboration between the DAOs. Establishing clear norms such as fair play, open communication, and mutual benefit is critical for such cases.

\smallskip
\noindent\textbf{Programming tools.} 
There are 4 publications mentioning programming tools for developments in DAOs. In addition to the automated development workflow that implements the typical DAOstack~\cite{daostack}, the considered programming tools mainly include anomaly detection to improve the security of smart contracts. DAO communities should spare efforts to establish security protocols for auditing code and develop improved DAO tooling or supportive infrastructure.

\smallskip
\noindent\textbf{Privacy-preserving approaches.}
Researchers also take interest in preserving privacy in DAOs. As voting mechanisms play a vital role in DAOs, anonymous voting desires increasing attention with several prominent approaches being considered such as the ring signature~\cite{ring} and zero-knowledge proof~\cite{zero}.



\begin{center}
\tcbset{
        enhanced,
        boxrule=0.4pt,
        fonttitle=\bfseries
       }
\begin{tcolorbox}[colback=gray!10,
colframe=black, 
width=3.4in,
arc = 1mm,
boxrule=0.5 pt,
lifted shadow={1mm}{-2mm}{3mm}{0.1mm}{black!50!white}
]
\footnotesize
\textbf{Findings of RQ-3: \textit{What techniques will likely be used in future DAO applications?}}
\vspace{0.2em}

\noindent\hangindent 0.6em$\triangleright$\,\textbf{Well-organized and collaborative DAOs.} The increasingly complex interaction and collaboration between different DAOs desire attention. Currently, the two most prominent approaches are DAO collaboration which helps to formalize the structure and communication of DAOs for higher efficiency, as well as SubDAOs which indicate hierarchical relationships across different DAOs for more concise management.

\noindent\hangindent 0.6em$\triangleright$\,\textbf{Security\&Privacy-improved approaches.}
Making use of anomaly detection tools for smart contracts in DAOs becomes the future. Anonymous voting is increasingly important as DAO participators are keen not to disclose who votes or the voting tally in order for a stronger privacy level.

\end{tcolorbox}
\end{center}

\label{sec_results}

\section{Conclusion}

In this work, we conduct a literature review, which is an extensive investigation of DAO-related studies, and analyze their corresponding features. Our results give clear answers to a series of hot topics on DAO's easy-understanding definition, architectural design, potential opportunities as well as to-be-improved challenges. From our view, we provide the first architectural-level exploration of DAO's development in the context of Web3 applications.
We here wish the following studies could extend the scope of literature review singly from academia to a wide range of in-the-wild DAO projects (11,000+, recorded on prevalent DAO launchpads, e.g., Snapshot, Tally, etc.), and combine the results with existing work to thoroughly investigate to what extent the community has leveraged the recommended architectural solutions for DAOs.

\bibliographystyle{IEEEtran}
\bibliography{bib(short).bib}

\appendix

\nociteappendix{*}

\bibliographystyleappendix{IEEEtran}
\bibliographyappendix{Appendix.bib}

\end{document}